\documentclass[twocolumn]{aastex62}

\usepackage{amsmath}
\usepackage{longtable}
\usepackage{natbib}

\submitjournal{AAS Journals}
\shorttitle{Asymmetric Fe in WASP-76b}
\shortauthors{Kesseli et al.}
\pdfoutput=1

\begin{document}

\title{Confirmation of Asymmetric Iron Absorption in WASP-76b with HARPS}

\correspondingauthor{Aurora Y. Kesseli}
\email{kesseli@strw.leidenuniv.nl}
\author[0000-0002-3239-5989]{Aurora Y. Kesseli}
\affiliation{Leiden Observatory, Leiden University, Postbus 9513, 2300 RA, Leiden, The Netherlands}

\author{I.A.G. Snellen}
\affiliation{Leiden Observatory, Leiden University, Postbus 9513, 2300 RA, Leiden, The Netherlands}

\begin{abstract}

Hot Jupiters are predicted to have hot, clear daysides and cooler, cloudy nightsides. Recently, an asymmetric signature of iron absorption has been resolved in the transmission spectrum of WASP-76b using ESPRESSO on ESO's Very large Telescope. This feature is interpreted as being due to condensation of iron on the nightside, resulting in a different absorption signature from the evening than from the morning limb of the planet. It represents the first time that a chemical gradient has been observed across the surface of a single exoplanet. In this work, we confirm the presence of the asymmetric iron feature using archival HARPS data of four transits. The detection shows that such features can also be resolved by observing multiple transits on smaller telescopes.  By increasing the number of planets where these condensation features are detected, we can make chemical comparisons between exoplanets and map condensation across a range of parameters for the first time. 

\vspace{25pt}

\end{abstract}

\section{Introduction}

In recent years, high-dispersion spectroscopy has transformed the field of exoplanet atmospheric characterization by resolving signatures of atmospheric dynamics in hot Jupiters. \citet{Snellen2010} first observed a possible net blueshift of CO absorption on HD 209458b and attributed it to a global wind moving from the hot dayside to the cooler nightside of the planet. These high-velocity winds were predicted by 3D global circulation models and are thought to homogenize the temperatures on the day and nightsides on these strongly irradiated gas giants \citep{Showman2008}. Since the first detection, global winds have been detected on many hot Jupiters with velocities up to 10 km s$^{-1}$ \citep[e.g.,][]{Alonso2019b, Casasayas2019, Nugroho2020}.

\citet{Ehrenreich2020} further advanced the use of high-dispersion spectroscopy to study atmospheric dynamics by resolving an asymmetric absorption signature in two individual transits of WASP-76b \citep{West2016}. This was possible for the first time thanks to the radial velocity precision of ESPRESSO on ESO's 8m Very Large Telescope (see also \citealt{Tabernero2020} for a subsequent analysis of the dataset, cataloging many atoms). \citet{Ehrenreich2020} find that while at the start of the transit an iron absorption signal is visible near the planet's rest frame velocity, at the end of the transit it is significantly blueshifted to $-11$ km s$^{-1}$. The cause of this signature was explained by a combination of planetary rotation and a global wind coming from the dayside, 
necessitating that the iron condenses out on the planet nightside. During a transit, the angle under which the planet is observed changes by almost 30$^\circ$, meaning that the irradiated hemisphere on the morningside is visible early on during the transit, while the eveningside is only seen during the second half of the transit. Assuming that there is no or very little iron absorption from the nightside of the planet due to rainout, the components of iron absorption from the morning and eveningside  rotate into and out of view, showing different levels of radial velocity shift due to the rotation and global wind counteracting and reinforcing each other respectively \citep{Ehrenreich2020}. This is the first time that a chemical gradient has been directly observed across the surface of an exoplanet, and demonstrates the future potential of high-dispersion spectroscopy to observationally constrain condensation and 3D circulation modeling.

Evidence for a similar asymmetry in iron has been seen in WASP-121b and MASCARA-2b as well, which further solidifies the presence of condensation signatures in many ultra-hot Jupiters. \citet{Bourrier2020} find a tentative offset in WASP-121b's systemic velocity derived from the entire cross correlation mask (which is dominated by iron lines) from the beginning to the end of the transit. Using ESPRESSO, \citet{Borsa2021} validated the presence of the blueshift variations in the iron signal during the transit of WASP-121b, finding a blueshift of $-2.80\pm0.28$ km s$^{-1}$ at the beginning of the transit and $-7.66\pm0.16$ km s$^{-1}$ at the end. Finally, \citet{Hoeijmakers2020b} find evidence of a strengthening absorption signature from iron in MASCARA-2b from the beginning to the end of the transit, which could also indicate changing chemistry across the surface of the planet. 

In this paper, we aim to confirm the detection of the iron signature in WASP-76b  using HARPS archival data. In Section \ref{s:data} we present the data, and in Section \ref{s:model} we explain how we created the model iron spectrum. In Section \ref{s:cc} we describe the cross correlation analysis between the data and the model. Next, we present the results of the iron condensation feature in Section \ref{s:results}. Finally, we summarize our conclusions in Section \ref{s:conclusions}. 

\begin{center}
\begin{table}[ht]
\caption{WASP-76 system parameters} 
\begin{tabular}{l l l}
\hline
Parameter & Value & Reference \\ 
\hline
\textbf{Star} & & \\
\hline
Spectral Type & F7 & W16$^1$\\
V-mag & 9.52$\pm0.03$ & CDS Simbad \\
$M_*$ ($M_{Sun}$) & $1.458\pm0.021$ & E20$^2$ \\
$R_*$ ($R_{Sun}$) & $1.756\pm0.071$ & E20 \\
$T_\mathrm{eff}$ (star; K) & 6250$\pm$100 & W16\\
$v_{sys}$ (km s$^{-1}$) & -1.16 & E20 \\
$v \sin i$ (km s$^{-1}$) & $1.48\pm0.28$ & E20 \\
$K_*$ (m s$^{-1}$) & $116.02 \substack{+1.29 \\-1.35}$ & E20 \\
\hline
\textbf{Planet} & & \\
\hline
$M_{p}$ ($M_{J}$) & $0.894\substack{+0.014 \\-0.013}$ & E20\\
$R_{p}$ ($R_{J}$) & $1.863\substack{+0.070\\-0.083}$  & E20 \\
log $g$ & 2.85 & calculated \\ 
$T_{eq}$ (K) & 2160$\pm$40 & W16 \\
$P_{orb}$ (d) & 1.809886 & W16 \\
$T_0$ (d) & 2456107.85507 & W16 \\
$a$ (au) & 0.0330$\pm$0.0005 & W16 \\
$K_p$ (km s$^{-1}$) & $196.52\pm0.94$ & E20 \\
$i$ (deg) & 88.0$\substack{+1.3\\-1.6}$ & W16 \\
$v_{rot}$ (km s$^{-1}$) & 5.1 & calculated \\
\hline
\footnotesize{
$^1$\citealt{West2016}}\\
\footnotesize{$^2$\citealt{Ehrenreich2020}}
\end{tabular}
\label{t:WASP76}
\end{table}
\end{center}

\vspace{-20pt}

\section{The Data}
\label{s:data}
WASP-76 is a bright F7-type main sequence star. The relevant parameters of the system are given in Table \ref{t:WASP76}. The data utilized in this work consist of time-series spectra taken during four different transits of WASP-76b.  All of the spectra were obtained with the High Accuracy Radial velocity Planet Searcher (HARPS; \citealt{Mayor2003}) at ESO's La Silla 3.6m telescope. HARPS has a spectral resolving power of $\sim$115,000 and covers a wavelength range from 378 to 691 nm. All of the data are publicly available and were downloaded from the ESO archive\footnote{\url{http://archive.eso.org/}}. Table \ref{t:obs} lists the observing details of the four separate nights. The transit on night 1 was published in \citet{Brown2017} and was used to derive an estimate of the Rossiter-McLaughlin effect. The transits on nights 2 and 3 were part of the Hot Exoplanet Atmospheres Resolved with Transit Spectroscopy (HEARTS) survey and were previously published in \citet{Seidel2019} as a focused study on the sodium doublet. Night 4 has not been previously published.

\begin{center}
\begin{deluxetable*}{lccccc}[!]
\tablecaption{Transit observations of WASP-76b with HARPS\label{t:obs}}
\tablehead{
\colhead{} &
\colhead{Date} & 
\colhead{PI \& ESO} & 
\colhead{Number of} & 
\colhead{Exposure Time} & 
\colhead{Average SNR} 
\vspace{-6pt}
\\
\colhead{} &
\colhead{} & 
\colhead{Prog. ID} &
\colhead{Spectra (N)} & 
\colhead{per Spec.} & 
\colhead{per Spec.} 
}
\startdata
Night 1 & 2012/11/11 &  Triaud (090.C-0540) & 61 & 300 s & 26.66 \\
Night 2 & 2017/10/24 & Ehrenreich (0100.C-0750) & 49 & $350 - 600$ s & 43.51\\
Night 3 & 2017/11/22 & Ehrenreich (0100.C-0750) & 65 & 300 s & 44.08\\
Night 4 & 2018/09/03 &  Louden, Wheatley \& Kirk (0101.C-0889) & 39 & 600 s & 45.77\\
\enddata
\end{deluxetable*}
\end{center}

\vspace{-30pt}

The data were processed with the HARPS Data Reduction Pipeline (DRS v3.5 for night 1 and v3.8 for nights 2, 3, and 4), which performs flat fielding, 2D extraction, and wavelength calibration. The pipeline produces 1D blaze-corrected, stitched spectra. The wavelength solutions have already been corrected for the barycentric velocity and are given in air. The spectra were subsequently shifted to the star's rest frame using the known system velocity ($v_{sys}$) and corrected for the reflex motion of the star due to the planet ($K_*$) at each observation. The spectra were then interpolated onto the same wavelength grid (of length 308100 pixels), masking any missing values at the wavelength extremes due to the different velocity shifts to create a 308100 by N wavelength grid in time for each night.  

Even though the blaze functions have been corrected by the pipeline, small differences in the slope of the stellar spectra due to imperfect blaze removal can cause problems when the stellar spectrum is removed in a later stage of the analysis. As we do not want to remove the shape of the spectrum as it contains information about the response function of the spectrograph, transmission through Earth's atmosphere and the signal from the host star, we followed a similar procedure as \citet{Hoeijmakers2019} and divided the 1D spectra into 50 parts, each of size $6162\times$N, and normalize each wavelength section to the mean flux in time. 

To remove any contamination from cosmic rays we performed a 3-sigma clipping procedure. We flagged any values that were more than 3-sigma different from the other time-series spectra at each wavelength pixel and interpolated over them in the wavelength direction. We also performed a masking of entire wavelength columns which had high standard deviations due to telluric lines, bad pixels, or deep stellar lines. We tested masking different percentages of pixels, and chose the value that most decreased the noise in the final cross correlation function without significantly altering the peak. This value ranged from about 5\% (nights 2, 3, and 4) to 15\% (night 1) of all pixels in the four different nights.  

\section{Model Spectra} 
\label{s:model}

We created a model transmission spectrum of iron using petitRADTRANS \citep{molliere2019}, in a method similar to \citet{Kesseli2020}.
petitRADTRANS is a radiative transfer code that creates emission or transmission spectra at low- or high-resolution. The high-resolution mode includes accurate line opacities for a large number of atoms, ions and molecules. We also tested using the same F9 radial velocity stellar mask as was used in \citet{Ehrenreich2020}, but found a stronger signal from the planet when we used a model atmosphere specifically tuned to the parameters of the planet (SNR of final detection increased from $\sim$5 to 9).

petitRADTRANS requires as input a temperature-pressure profile, the planet's surface gravity and radius, the abundances of the requested species, and the mean molecular weight (MMW) of the atmosphere. 
We used the planet radius and surface gravity that are listed in Table \ref{t:WASP76}, and an isothermal temperature-pressure profile at the equilibrium temperature of the planet. We experimented with different temperatures and found that changes of 150 K could change the height of the retrieved signal by small amounts ($\sim$50 ppm), but did not change the velocity position. We implemented an equilibrium chemistry code from \citet{Molliere2017} to determine the MMW of the planet at each altitude, and assumed a constant VMR of $10^{-5}$ for Fe at all altitudes. While a constant VMR at each altitude is not realistic, neither this assumption nor an order of magnitude change to the VMR significantly alter our recovered planet signal. A VMR of $10^{-5}$ is motivated by previous chemical modeling of hot Jupiters \citep{Lodders2002, Visscher2010, Hoeijmakers2020}. Finally, we added continuum opacity from H$_{2}-$H$_{2}$ collisions, H$_{2}-$He collision, and H$^{-}$.   

Finally, we reduced the resolution of the model to the resolution of the HARPS spectrograph by convolving it with a Gaussian kernel with a FWHM of 2.7 km s$^{-1}$. We did not perform any steps to take into account the planet's rotation, as the asymmetry in the iron signal is thought to be due to iron being present mostly on one limb of the planet. 

\section{Cross Correlation Analysis} 
\label{s:cc}

We followed the steps of \citet{Hoeijmakers2020} for the cross correlation analysis, and used the following equation 

\begin{equation} 
c(\nu, t) = \sum_{i=0}^{M} x_i(t) T_i(\nu) 
\end{equation}

\noindent where $c(\nu, t)$ is the resulting cross correlation grid for each radial velocity value ($\nu$) at each time ($t$). We define $x_i(t)$ as the observed spectrum at each time and $T_i(\nu)$ as the model at each velocity. The model was shifted from $-250$ to $+250$ km s$^{-1}$ in 1 km s$^{-1}$ steps, and then interpolated onto the same wavelength grid as the data. We normalized the model such that $\sum_{i=0}^{M} T_i(\nu) = 1$. In this way, the cross correlation function sums all of the pixels ($i$) in the spectrum, weighting them by their absorption.

\begin{figure*}
\begin{center}
\includegraphics[width= 0.48\linewidth]{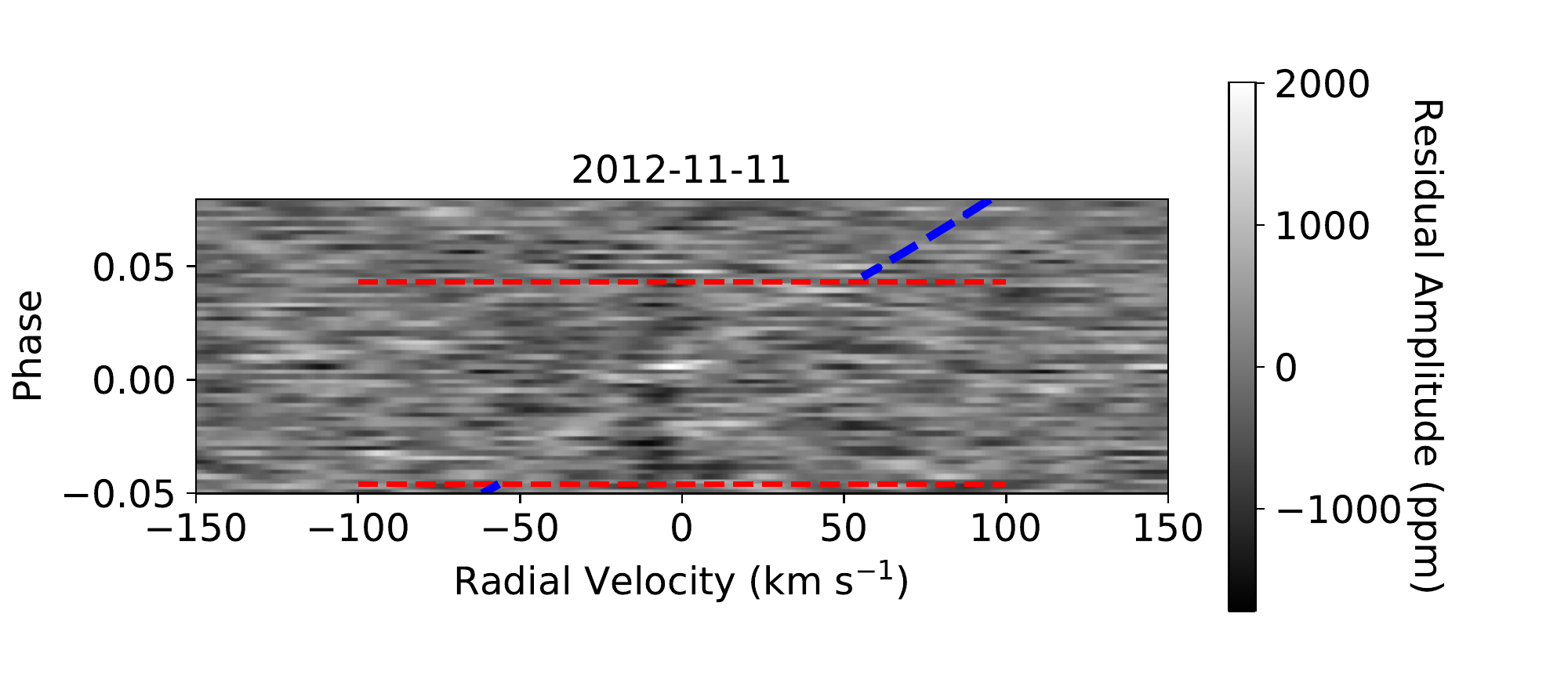}
\includegraphics[width= 0.48\linewidth]{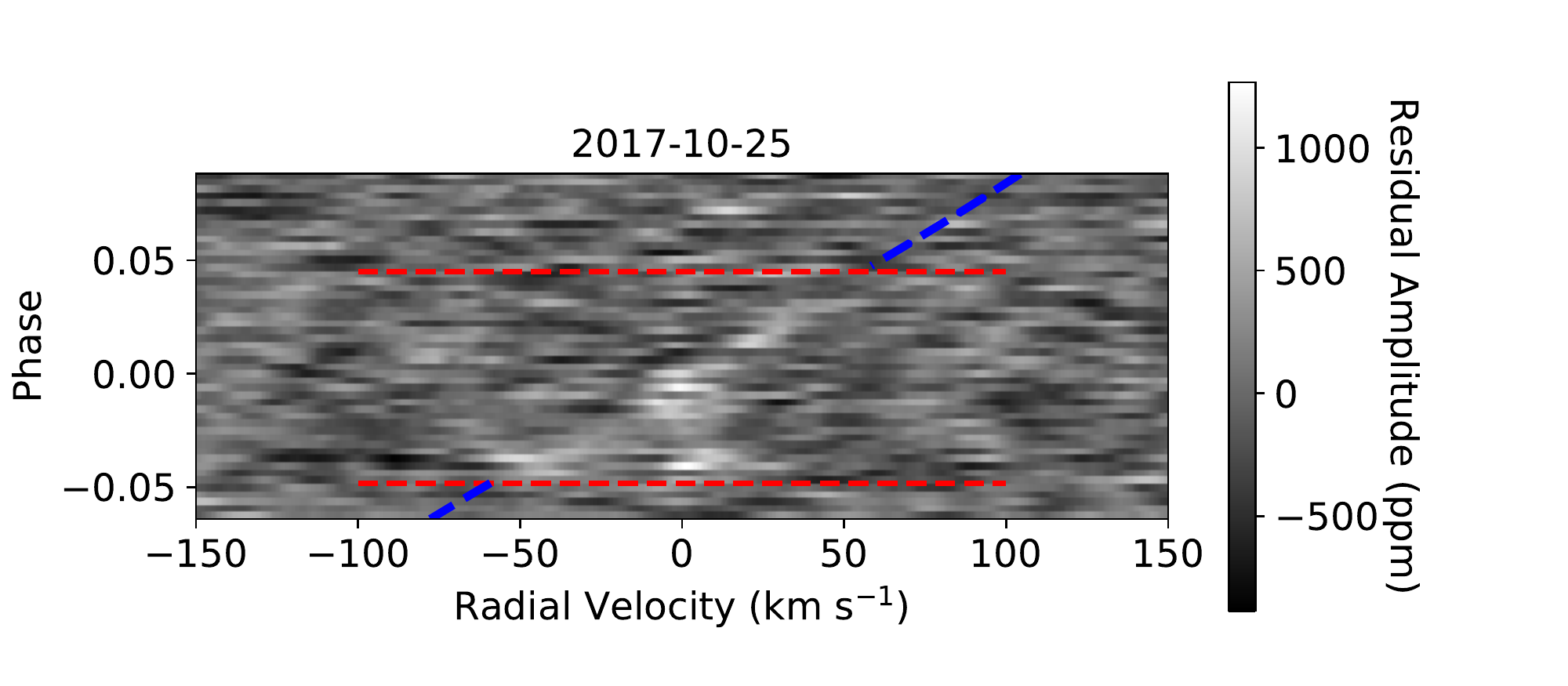}
\includegraphics[width= 0.48\linewidth]{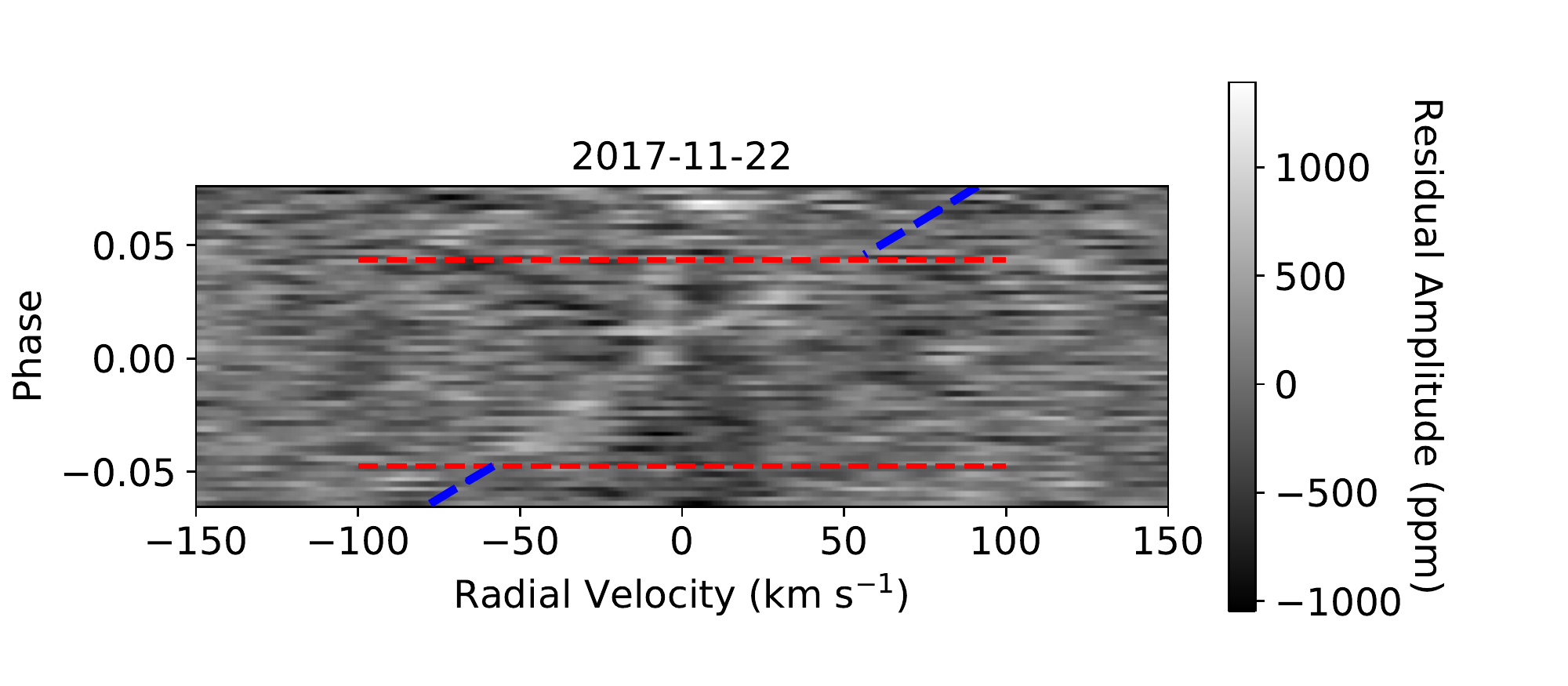}
\includegraphics[width= 0.48\linewidth]{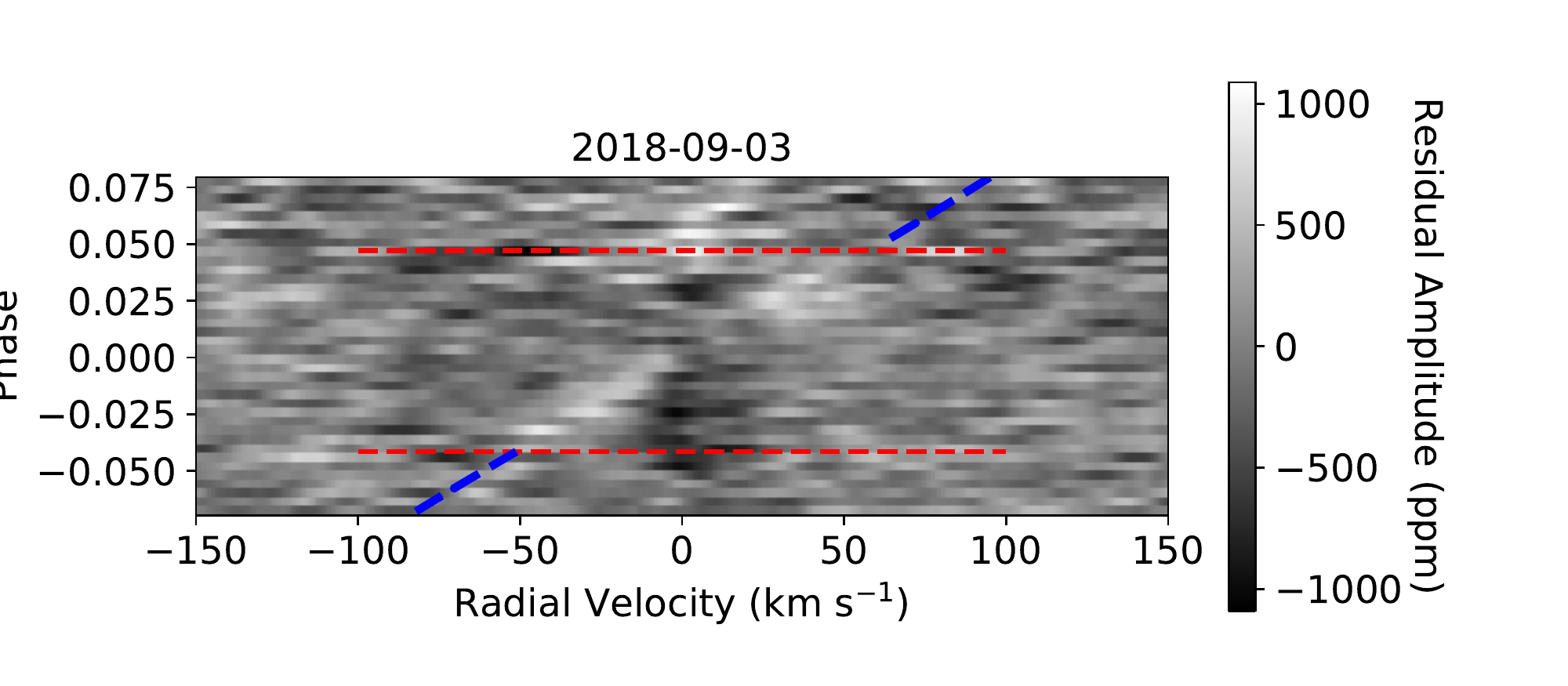}
\caption{\small 
Iron cross correlation grid for each of the four nights. The red lines show where the transits begin and end, while the blue lines show the expected path of the exoplanet. For each night the trace of the exoplanet is visible, but there is significant noise around 0 km s$^{-1}$ due to imperfect removal of stellar lines and the Doppler shadow. The residuals are largest, making the signal difficult to see, on Night 1 (2012-11-11), when the signal-to-noise ratio was on average 20 less than the other 3 nights.}
\label{f:cc_matrices}
\end{center}
\end{figure*}

At this point the contribution from the host star still needs to be removed, which dominates the cross correlation function. In order to remove the host star, we divided each cross correlation function by the average out-of-transit cross correlation function. After this step we were left with the residual amplitude (in ppm), which corresponds to the transit depth due to atmospheric Fe.

\begin{figure*}
\begin{center}
\includegraphics[width= 0.48\linewidth]{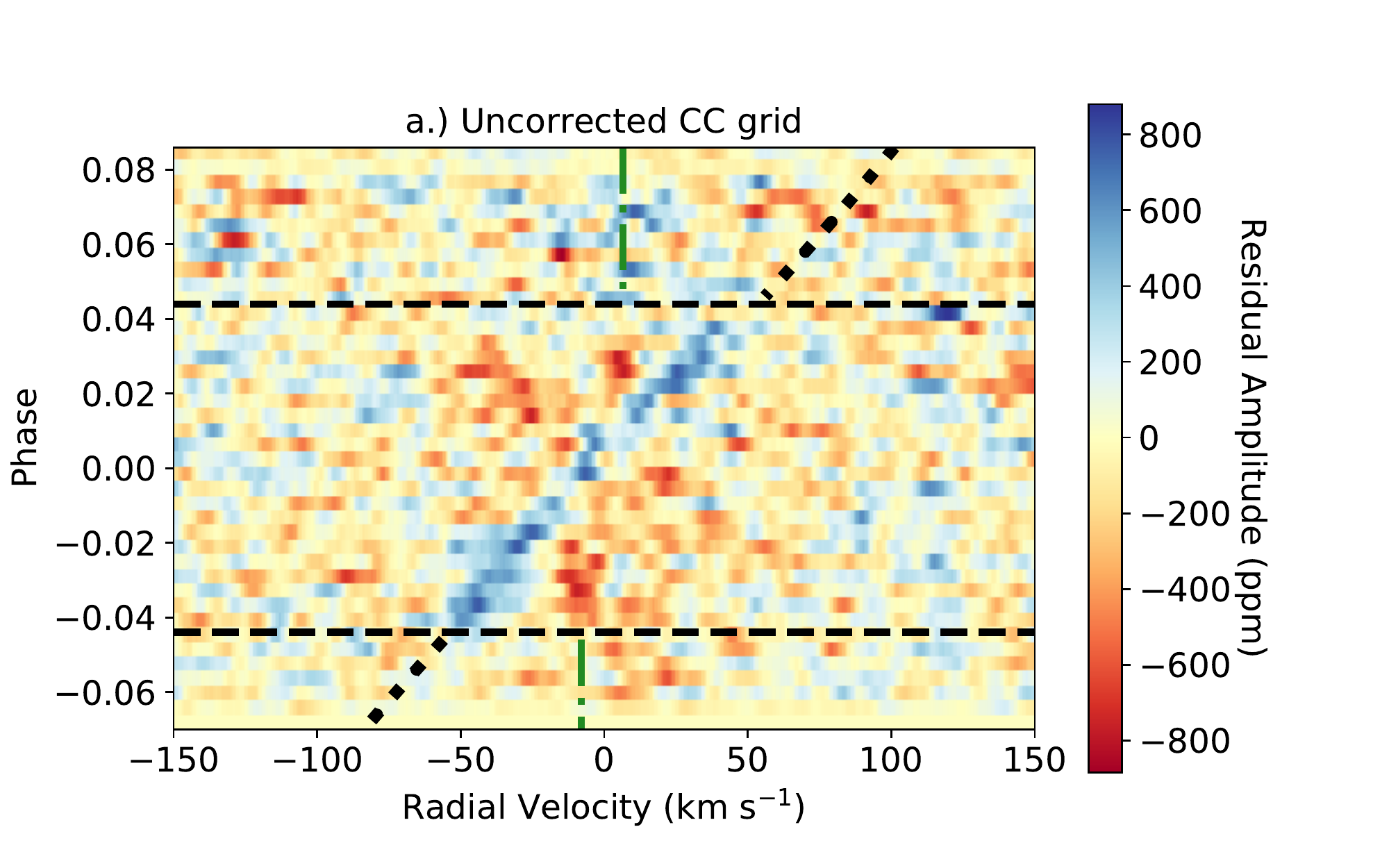}
\includegraphics[width= 0.48\linewidth]{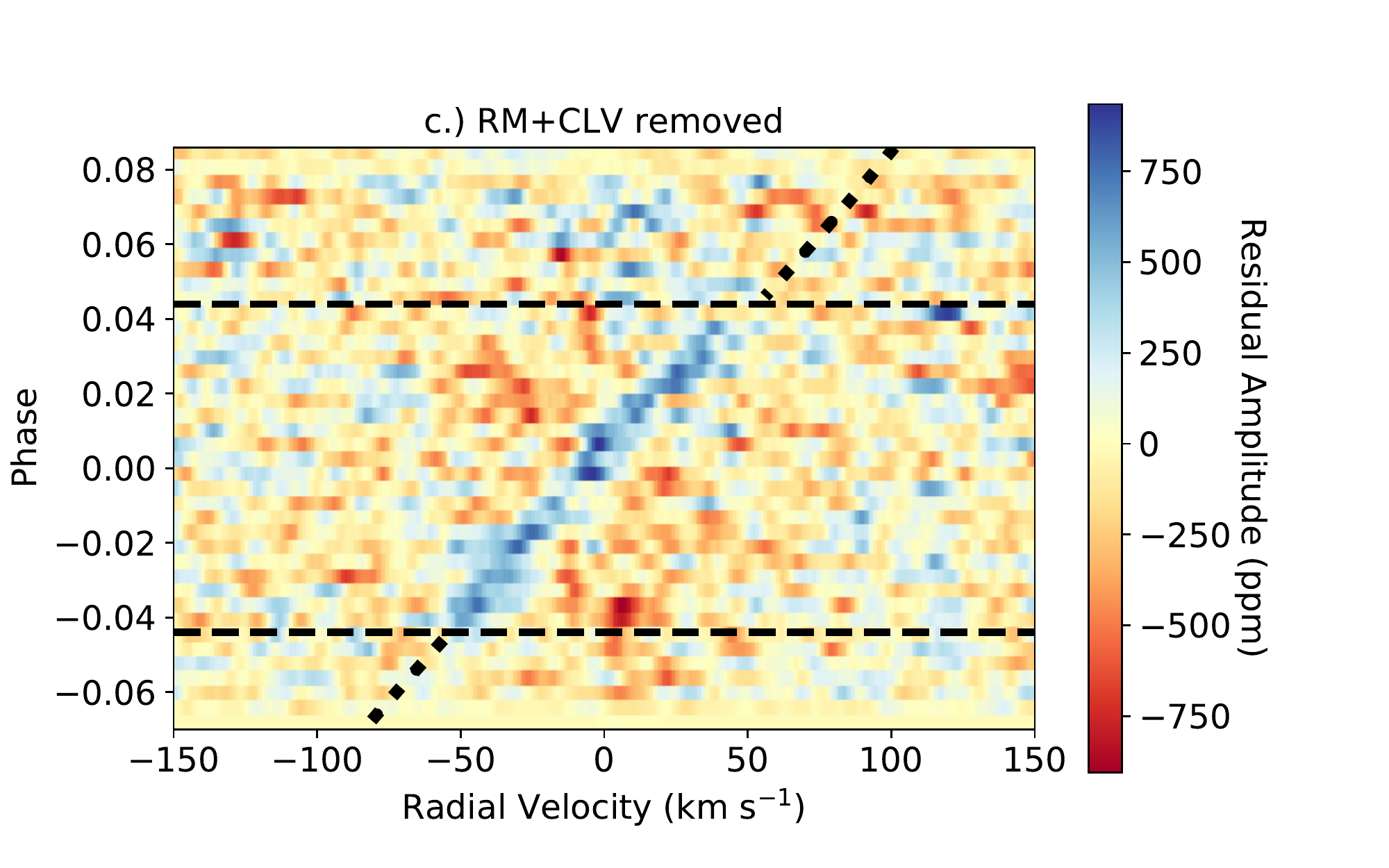}
\includegraphics[width= 0.48\linewidth]{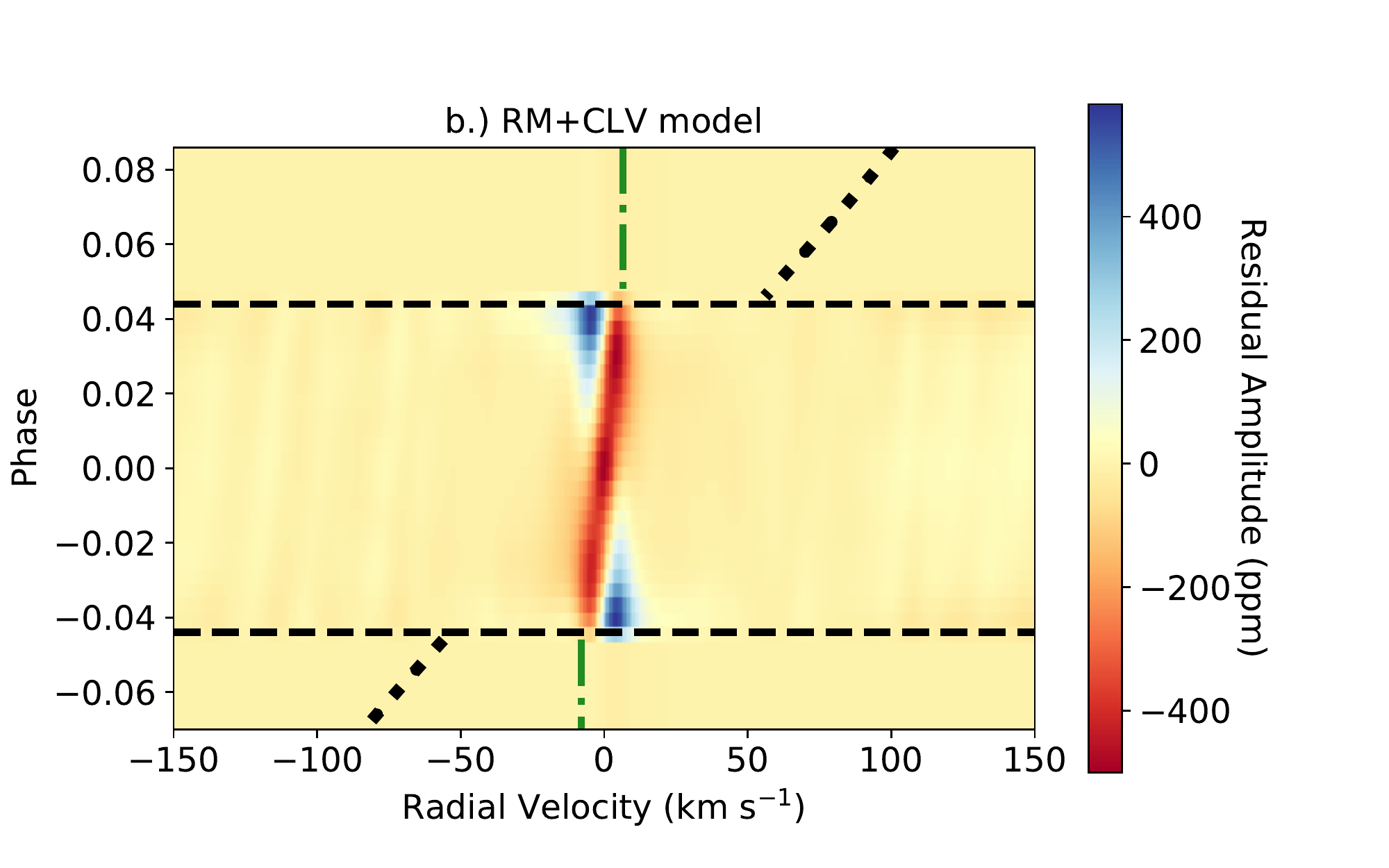}
\includegraphics[width= 0.48\linewidth]{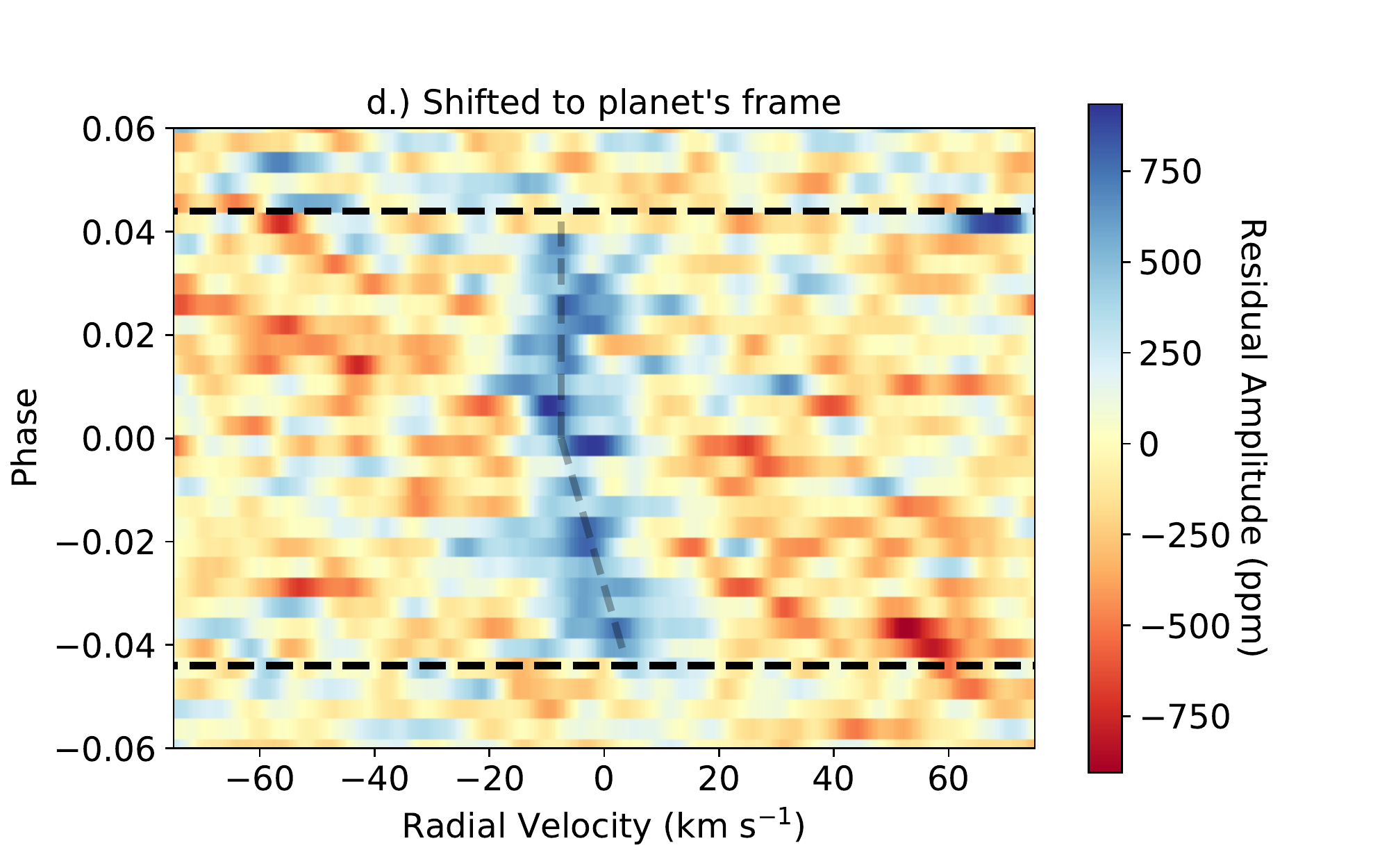}
\caption{\small 
\textbf{a.)} Combined cross correlation grid in the rest frame of the star. The exoplanet's trace is clearly visible, as well as a small residual signal due to the Doppler shadow (green dashed line). \textbf{b.)}  Model of the Rossiter McLaughlin effect and the center to limb variations. \textbf{c.)} 2D combined cross correlation grid, where the Doppler shadow and center-to-limb variations have been removed by subtracting our model (panel b). \textbf{d.)} The cross correlation grid shifted to the exoplanet rest frame. Without winds or any atmospheric dynamics, the signal should lie vertically at 0 km s$^{-1}$. However, around phases of 0.0 to +0.04 the peak in the cross correlation function is shifted to about -8 km s$^{-1}$, while at the beginning of the transit (phase of -0.04) the peak signal is closer to 0 km s$^{-1}$. }
\label{f:cc_comb}
\end{center}
\end{figure*}

The cross correlation matrices for each night are shown in Figure \ref{f:cc_matrices}. Each night has an excess along the exoplanet's expected velocity, but in all of the individual nights there are significant residuals around 0 km s$^{-1}$ due to a combination of the Doppler shadow and imperfect removal of stellar lines. By combining all of the transits together, we reduced much of this noise and were able to more easily determine the shape of the trace. To combine the nights together, we interpolated the cross correlation grids onto a new cross correlation grid that was uniformly sampled along the phase axis in steps of 0.004. We chose this step size to be slightly larger than the average phase step size of each night, which ranged from about 0.002 to 0.0038, so that none of the values would be extrapolated. We then averaged the four nights together, weighting each by its average signal-to-noise ratio and the number of spectra during the transit (Table \ref{t:obs}). The left panel of Figure \ref{f:cc_comb} shows the averaged cross correlation matrix from all four nights. The exoplanet's trace can now clearly be seen.

\subsection{The Doppler Shadow}

Due to the orientation of the planet-star system \citep[see][]{Ehrenreich2020}, the Doppler shadow effect is less than a HARPS radial velocity pixel (2.7 km s$^{-1}$) and so it is difficult to distinguish in the data sets individually. Our use of a model atmosphere tuned to the parameters of WASP-76b as opposed to an stellar binary mask, as was done in \citet{Ehrenreich2020}, also led to a much less pronounced Doppler shadow than was seen in that work. However, a weak Doppler shadow can be seen in the top left panel of Figure \ref{f:cc_comb} as a red residual near the green dashed lines. In addition to the Doppler shadow signature, residuals due to center-to-limb variations (CLVs) of the stellar spectrum are also expected. 

We removed the Doppler shadow and the center-to-limb variations in a manner similar to \citet{Yan2018} and \citet{Casasayas2019}. CLVs can be resolved on the surface of the Sun, and because WASP-76 has an effective temperature within 8\% of the effective temperature of the Sun, we used high resolution observations of the Sun at different limb darkening angles from \citet{Stenflo2015} to estimate the CLVs. The spectra include 10 different limb-darkening angles of $\mu = 0.1$ to 1.0. The wavelength coverage of the Solar observations do not extend quite as far into the blue as the HARPS spectra and the bluest 30 nm are missing. This loss is less than 10\% of the spectrum, and so our CLV effect estimates could be slightly underestimated. Because of this imperfect match, we modeled the Rossiter McLaughlin effect with a BT-Settl model atmosphere \citep{Allard2012} at the temperature and surface gravity of WASP-76.

To model the transit of WASP-76b over the surface of WASP-76, we divided the surface of the star in $0.01 \times 0.01 R_*$ pieces and assigned each one a velocity based on the star's known $v \sin i$ and a limb darkening angle, $\mu$. At each phase we mapped the exoplanet's path across the surface of the star and combined all the stellar spectra at different limb darkening angles and velocities except those covered by the planet. This produced mock observations at each phase, one for the CLV effect and one for the Doppler shadow. 

We then performed the same cross correlation analysis with the modeled observations as we did with the real data. We combined the Doppler shadow model and the CLV model by simply adding them together. The resulting model is shown in the bottom left side of Figure \ref{f:cc_comb}. To remove the Doppler shadow and CLV effects, we subtracted the model from the cross correlation plot. At this point we also applied a high pass filter with a width of 70 km s$^{-1}$ to remove any broadband structure left from imperfect blaze function removal. The final combined and corrected cross correlation function is shown in the top right side of Figure \ref{f:cc_comb}.

\section{Results} 
\label{s:results}

To determine whether the shape of the iron absorption is indeed asymmetric during the transit, we shifted the exoplanet into it's expected rest frame using the value of $K_p$ from Table \ref{t:WASP76} (bottom right panel of Figure \ref{f:cc_comb}). Between phases of 0.0 and +0.04 (the second half of the transit) all of the cross correlation functions peak around $-8$ km s$^{-1}$, except for two pixel at a phase of 0.028 and 0.022. Before a phase of $-0.01$ the peaks shift towards 0 km s$^{-1}$, and at the very beginning of the transit (phase = $-0.04$), the peak of the cross correlation function is at +1.5 km s$^{-1}$. 

\begin{figure*}
\begin{center}
\includegraphics[width=0.48\linewidth]{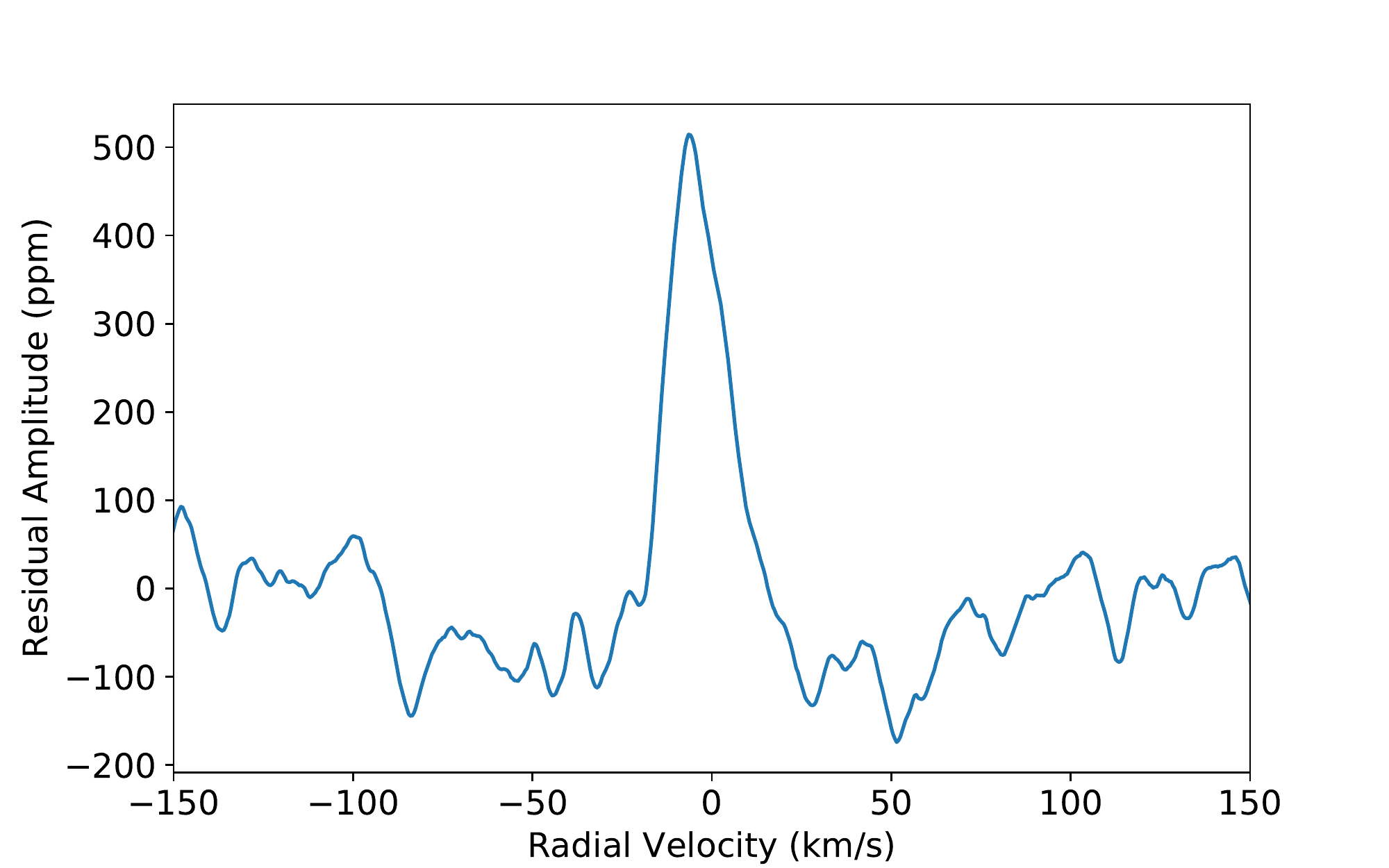}
\includegraphics[width=0.48\linewidth]{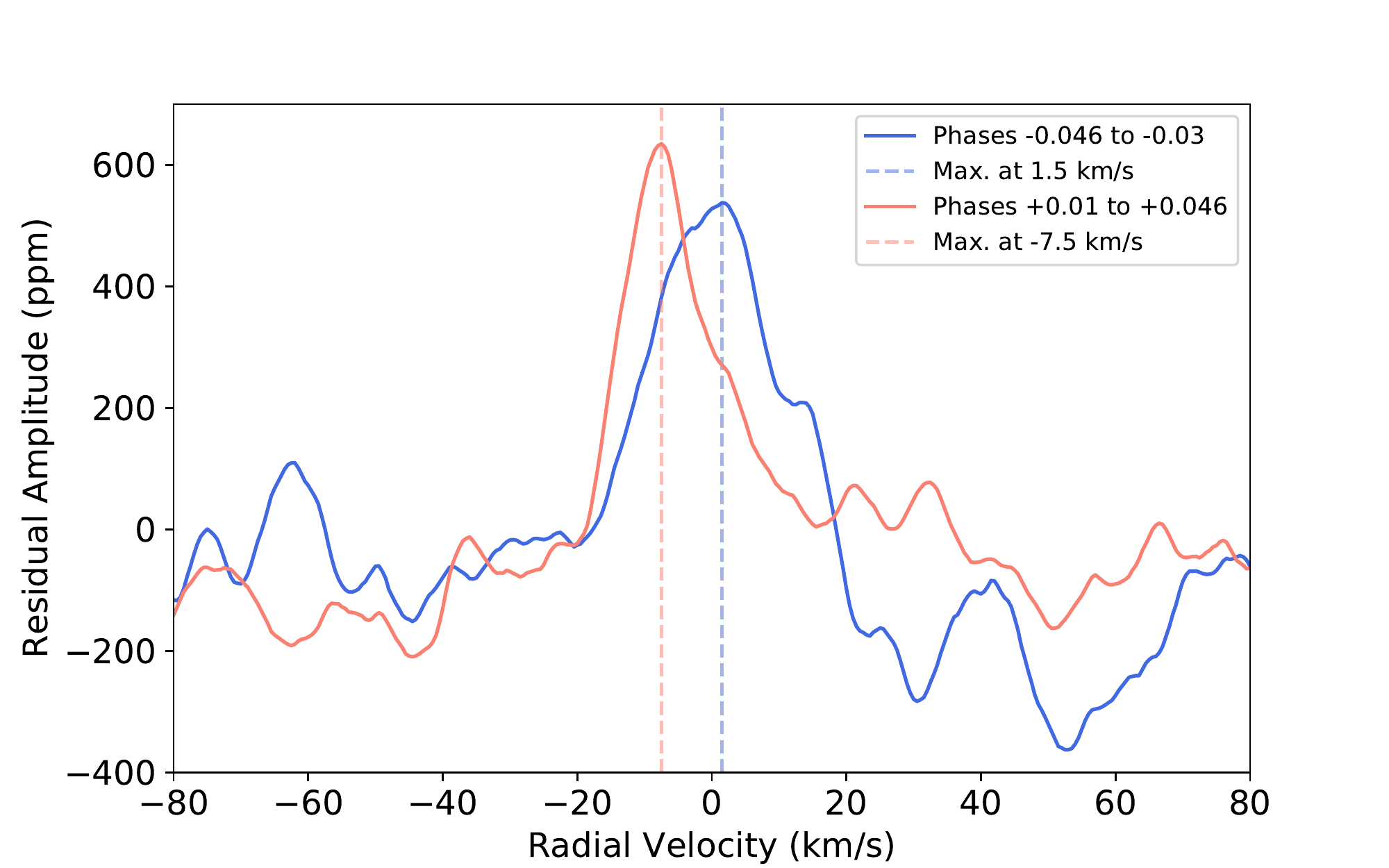}
\caption{\small 
\textbf{Left}: 1D cross correlation function of iron created by averaging together all of the in-transit cross correlation functions from panel d of Figure \ref{f:cc_comb}. We find that the amplitude of the Fe signal is about 500 ppm and that Fe is detected with a SNR of 9.22. \textbf{Right}: the 2D cross correlation grid was split into two different phase bins and the cross correlation functions in these bins were averaged together. At the beginning of the transit the the signal reaches a maximum at 1.5 km s$^{-1}$, while during the second half of the transit the cross correlation signal reaches a maximum at $-7.5$ km s$^{-1}$. Similar to \citet{Ehrenreich2020} we find that at the end of the transit the signal is stronger.  }
\label{f:cc_flat}
\end{center}
\end{figure*}

We combined the two dimensional cross correlation function grid into a single cross correlation function by averaging the phases during the transit together (Figure \ref{f:cc_flat}). The peak residual amplitude of the feature is at about 500 ppm. If we average the beginning of the transit together and the end of the transit together separately, we find that the beginning of the transit has a peak radial velocity of $+1.5$ km s$^{-1}$, while the end of the transit has a peak at $-7.5$ km s$^{-1}$. The velocities of this asymmetry are similar to the ones found in \citet{Ehrenreich2020}.  

We calculated uncertainties on these radial velocity shifts by fitting Gaussians to the 1D cross correlation function of the beginning of the transit and the end of the transit separately (red and blue curves shown in the right panel of Figure \ref{f:cc_flat}). The Gaussian for the beginning of the transit has a center at $+0.04\pm 1.47$ km s$^{-1}$, while the Gaussian for the end of the transit has a center at $-6.9 \pm 0.75$ km s$^{-1}$. The stated uncertainties in the measurement are calculated by dividing the standard deviation of the Gaussian by the signal to noise ratio of the line. As a secondary test, we also checked these values by taking the average and standard deviation of the individual values in each each night at the beginning and end of the transits. For each night, we recorded the value of any peak present between $-25$ and $+10$ km s$^{-1}$ for each cross correlation function between phases $+0.042$ and $+0.002$ for the second half of the transit and $-0.026$ and $-0.042$ for the first half of the transit. The average radial velocity in the first half of the transit was $+0.3\pm1.07$, while the average radial velocity in the second half of the transit was $-7.1 \pm 0.74$ km s$^{-1}$. The uncertainties in the peak are given by the standard deviation divided by the square root of the number of measurements. Both methods return very similar results and show that the peaks are not consistent within two standard deviation, and so we consider this a confirmation of the asymmetric iron absorption feature.

We find that the cross correlation peak during the second half of the transit is stronger than the peak during the beginning of the transit by about 100 ppm. This is consistent with the results from \citet{Ehrenreich2020} and aids in the hypothesis that the asymmetry arises from more iron being present on the evening side of the planet than the morning side, where monatomic iron has condensed out of the atmosphere. The iron signal from the four nights of HARPS data presented here, and the iron signal presented in \citet{Ehrenreich2020} both have noise levels of about 50 ppm per pixel. By taking into account this noise level, the wavelength steps and resolutions of the two spectrographs, the signal amplitude, and assuming unresolved iron lines, we find that 7.3 HARPS transits are required to achieve similar results to one ESPRESSO transit.

\section{Conclusions} 
\label{s:conclusions}

We analyzed four publicly available transits of WASP-76b from the HARPS instrument on ESO's La Silla 3.6m telescope. To search for atomic absorption from Fe, we created a model of the transmission spectrum using petitRADTRANS. We then cross correlated the model with the data using a method similar to \citet{Hoeijmakers2020}. We resolved a small Doppler shadow signature, and removed it by modeling the stellar spectra at each phase. 

We found clear evidence of absorption from Fe at a SNR of $\sim$9, and find an average residual amplitude of 500 ppm across the transit. Our analysis shows an asymmetric velocity shift from the morning to the evening side of the planet. We divided the transit into phase bins and determine the radial velocity of the peak and uncertainty of this peak for the beginning and end of the transit separately. At the beginning of the transit, the signal is best fit by a Gaussian with a peak of $+0.04\pm 1.47$ km s$^{-1}$, while during the second half of the transit the peak is at $-6.9 \pm 0.75$ km s$^{-1}$. The morning and evening radial velocities that we find are not within two standard deviation of each other, further confirming the results from \citet{Ehrenreich2020}.

Additional chemical modeling and observations are required to determine the process that is removing iron from the gas phase on the nightside of WASP-76b (i.e. condensation into metallic iron or other iron molecules). FeH is predicted to be the second most abundant iron carrier at this temperature after monatomic Fe \citep{Visscher2010}, and so some Fe gas could condense into FeH gas, but \citet{Kesseli2020} did not find any evidence of FeH in WASP-76b. More observations of Fe condensation and other Fe molecules on a range of exoplanets will aid in the physical interpretation of this signature. 

\acknowledgements

A.K. and I.S. acknowledge funding from the European Research Council (ERC) under the European Union's Horizon 2020 research and innovation program under grant agreement No 694513. This paper is based on data obtained from the ESO Science Archive Facility under request number 569981 by Aurora Kesseli. This research has made use of the NASA Exoplanet Archive, which is operated by the California Institute of Technology, under contract with the National Aeronautics and Space Administration under the Exoplanet Exploration Program. 

\facilities{ESO:3.6m (HARPS)}

\software{\texttt{astropy} \citep{astropy}, \texttt{matplotlib} \citep{matplotlib}, \texttt{numpy} \citep{numpy}}

\bibliographystyle{aasjournal}
\bibliography{bib.bib}

\end{document}